\documentclass[conference]{IEEEtran}
\usepackage{cite}
\ifCLASSINFOpdf
  \usepackage[pdftex]{graphicx}
\else
  \usepackage[dvips]{graphicx}
\fi
\usepackage{url}
\usepackage[utf8]{inputenc}
\usepackage[T1]{fontenc}
\begin{document}
\title{PRACH Preamble Detection as a Multi-Class Classification Problem: A Machine Learning Approach Using SVM}

\author{\IEEEauthorblockN{Filip Ferenc\textsuperscript{1,2}, Michał Szcząchor\textsuperscript{2}}
\IEEEauthorblockA{\textsuperscript{1}Wrocław University of Science and Technology\\
27 Wybrzeże S. Wyspiańskiego\\
50-370 Wrocław, Poland}\\
\IEEEauthorblockA{\textsuperscript{2}Nokia\\
Szybowcowa 2\\
54-130 Wrocław, Poland\\
Email: filip.ferenc@nokia.com, michal.szczachor@nokia.com}}

% \and
% \IEEEauthorblockN{James Kirk\\ and Montgomery Scott}
% \IEEEauthorblockA{Starfleet Academy\\
% San Francisco, California 96678-2391\\
% Telephone: (800) 555--1212\\


%Telephone: (800) 555--1212, Fax: (888) 555--1212}


\maketitle

\begin{abstract}
This study addresses the preamble detection problem in the Random Access procedure of LTE/5G networks by formulating it as a multi-class classification task and evaluating the effectiveness of machine learning techniques. A Support Vector Machine (SVM) model is implemented and compared against conventional detection methods. The proposed approach improves preamble index assignment, enhancing detection efficiency for User Equipment (UE) accessing the network. Performance analysis demonstrates that the SVM-based solution increases detection accuracy while reducing missed detections. These findings underscore the potential of machine learning in optimizing the Random Access procedure and improving network accessibility.
\end{abstract}


% \fi
\IEEEpeerreviewmaketitle

\section{Introduction}
In LTE/5G networks, efficient preamble detection in the Random Access (RA) procedure \cite{3GPP-TR-38.802,3GPP-TR-38.801} is essential for ensuring reliable network access for User Equipment (UE). The RA procedure enables UEs to establish a connection with the base station, but its efficiency is highly dependent on accurate preamble detection. Traditional detection methods often struggle with false alarms and missed detections, which can degrade network performance and accessibility.

To overcome these challenges, this study redefines the preamble detection problem at the base station as a multi-class classification task and investigates the potential of machine learning techniques. The work on the application of ML algorithms in the task of PRACH signal detection has already been considered in \cite{Guel2024,Modina2019,Singh2024,Zehra2022,Li2024}, however, this work was focused on neural networks. In this paper, we present a different approach to the problem using the Support Vector Machine algorithm (SVM). A SVM model is developed and assessed in comparison to traditional detection methods. By utilizing machine learning, the proposed approach seeks to optimize preamble index assignment, enhancing detection efficiency and minimizing errors in network access attempts.

To minimize the number of input parameters and identify the rate that highlights the signal features, Principal Component Analysis (PCA) decomposition was employed. As a result, the principal component space proved to be more discriminative compared to the original signal space spanned by the Zadoff-Chu (ZC) sequence \cite{ZC-seq}.

The final section of this article presents a comparison of analytical algorithms and the proposed machine learning solution across several propagation channel models, highlighting the potential capabilities of machine learning methods in preamble detection applications.

\section{Analytical Algorithm}

\begin{figure}[!h]
\centering\includegraphics[width=0.5\textwidth]{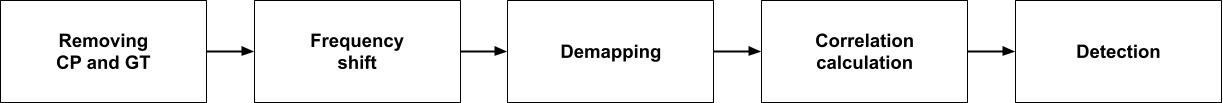}
\caption{Block diagram of the PRACH receiver.}
\label{fig:block_receiver}
\end{figure}

\begin{figure}[!h]
\centering\includegraphics[width=0.5\textwidth]{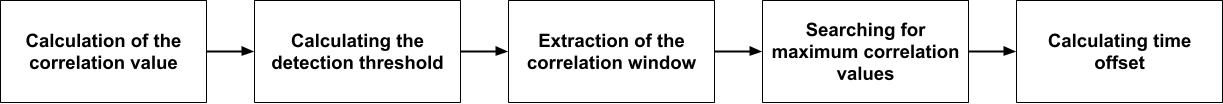}
\caption{Block diagram of the analytical preamble detector.}
\label{fig:block_algorithm}
\end{figure}

The conventional approach to PRACH \cite{MathWorksRACH,ShareTechNote} signal detection relies on an analytical algorithm, as described in \cite{Modina2019}, which follows a one-step detection process. However, in practical implementations, multi-stage detection algorithms are more commonly utilized due to their enhanced robustness and accuracy. The detection process of the PRACH signal begins with frequency synchronization between the Base Station (BS) and the User Equipment (UE). Following synchronization, the Cyclic Prefix and Guard Time are removed. The BS then configures the detection window to match the duration of the preamble  \cite{3GPP-TS-38.211} (Figure~\ref{fig:block_receiver} presents the block diagram of the PRACH receiver). Within this detection window, the BS performs a search for the highest correlation values by comparing the received signal with a predefined set of stored preamble patterns. The detection process is based on computing the cyclic correlation between the received signal and these preamble patterns. A successful detection occurs when the correlation value exceeds a predefined detection threshold.
The core component responsible for executing this detection mechanism is referred to as the detector, whose structure is illustrated in Figure~\ref{fig:block_algorithm}.

\section{Metrics for Evaluating Algorithms}

The evaluation of a PRACH receiver's performance \cite{3GPP-TS-38.101-4} is based on several fundamental metrics, primarily the False Alarm Rate (FA) and the Miss Detection Rate (MD), which are defined as follows:

\begin{itemize}
    \item \textbf{False Alarm Rate (FA)} – The proportion of detected preambles that either have an incorrect index compared to the transmitted preamble or are detected despite not being transmitted.
    \item \textbf{Miss Detection Rate (MD)} – The proportion of transmitted preambles that were not successfully detected.
\end{itemize}

Additionally, the effectiveness of the Support Vector Machine (SVM) algorithm was assessed using the following metrics:

\begin{itemize}
    \item \textbf{Validation Set Accuracy} – A measure of the model’s ability to correctly classify instances in a dataset that were not used during training.
    \item \textbf{Confusion Matrix} – A tabular representation of the classification algorithm’s performance, where each column corresponds to the actual class labels and each row represents the predicted labels.
\end{itemize}

The FA, MD, and confusion matrix metrics are interrelated. In evaluations based on machine learning algorithms, it is essential to analyze the entire confusion matrix to achieve a comprehensive assessment of the model’s performance.

\section{Database}

The dataset utilized for evaluating the performance of the analytical algorithm, as well as for training and testing the SVM-based model, was generated using an LTE/5G physical layer simulator provided by Nokia Solutions \& Networks. This simulator, implemented in Matlab, is highly configurable and enables comprehensive simulations of LTE/5G networks. It also facilitates the storage of selected variables during simulations and the generation of statistical metrics such as the False Alarm Rate (FA) and the Miss Detection Rate (MD).

The training dataset was generated under the following conditions:

\begin{itemize}
    \item The transmission Signal-to-Noise Ratio (SNR) varied from -20 dB to 20 dB,
    \item The simulation duration was set to 1 second,
    \item Two physical receiver antennas were used,
    \item A single User Equipment (UE) was present in the simulation,
    \item A fixed Timing Advance (TA) was applied,
    \item The PRACH sequence length was generated for the long format 0,
    \item Four propagation channel models were considered: Additive White Gaussian Noise (AWGN), Extended Pedestrian A (EPA), Extended Vehicular A (EVA), and Extended Typical Urban (ETU).
\end{itemize}

The dataset was generated for 64 preambles over a range of 41 SNR values. Each sample was collected during a one-second simulation, with a transmission window available every 10 ms \cite{keysight-5G,Lin2019-5G}, resulting in a total of 26,240 signal samples. Furthermore, as the data was recorded for two receiver antennas, each preamble at a given SNR was associated with 2 × 100 signal samples. After final processing, the complete dataset comprised approximately 2 million signal samples.

\section{Preprocessing}
The ZC complex signals, each consisting of 839 samples, were decomposed into their real and imaginary components.
\subsection{Data organization}
These components were subsequently arranged sequentially into a vector, resulting in a final vector representation consisting of 1678 samples.

\begin{figure}[!h]
\centering\includegraphics[width=0.5\textwidth]{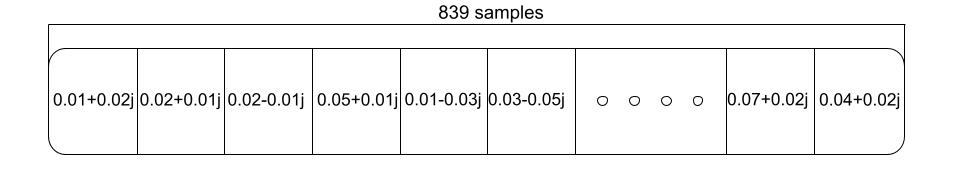}
\caption{Preamble vector prior to preprocessing.}  \label{fig:839pr}
\end{figure}

\begin{figure}[!h]
\centering\includegraphics[width=0.5\textwidth]{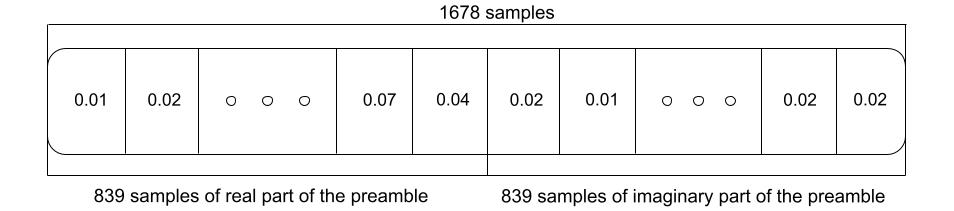}
\caption{Preamble vector following preprocessing.}  \label{fig:1678pr}
\end{figure}

Figures \ref{fig:839pr} and \ref{fig:1678pr} depict an example preamble vector before and after the preprocessing step.

Subsequently, the processed signals were organized into a matrix, where each row represented an individual signal and each column corresponded to a specific sample within that signal. Consequently, the columns of the matrix encapsulated distinct signal characteristics, treating each sample as an independent feature. This structured representation enabled comprehensive analysis through Principal Component Analysis (PCA) and facilitated signal classification using the Support Vector Machine (SVM) algorithm.

\begin{figure}[!h]
\centering\includegraphics[width=0.5\textwidth]{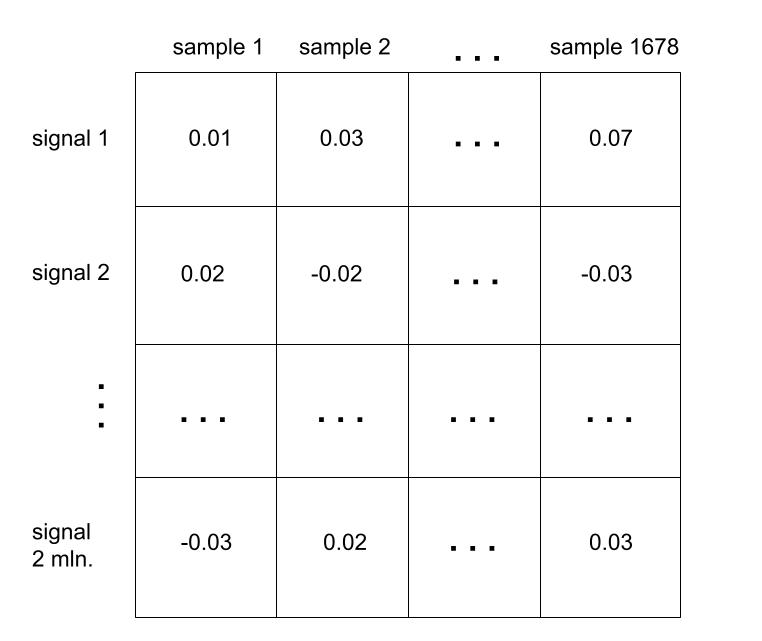}
\caption{Signal matrix after preprocessing.}  \label{rys:matrix_sig}
\end{figure}

Figure \ref{rys:matrix_sig} illustrates the structure of the signal matrix generated according to the described approach.

\subsection{Principal Component Analysis}
Principal Component Analysis (PCA) \cite{Jolliffe-Cadima-PCA,MathWorksPCA} is a statistical method that utilizes the decomposition \cite{eriksen2023data} of a data matrix to determine the directions along which the variance of the data is maximized. This technique transforms the coordinate system such that the first principal component represents the axis along which the data exhibits the greatest variance. Subsequent principal components are mutually orthogonal and are arranged in descending order of variance contribution.

The findings of the PCA analysis are presented below. The figures illustrate the cumulative variance as a function of the number of components, as well as a visualization of the data projected onto the first two principal components.
\begin{figure}[ht]
	\centering
	\begin{minipage}[b]{0.49\textwidth}
		\centering\includegraphics[width=0.98\textwidth]{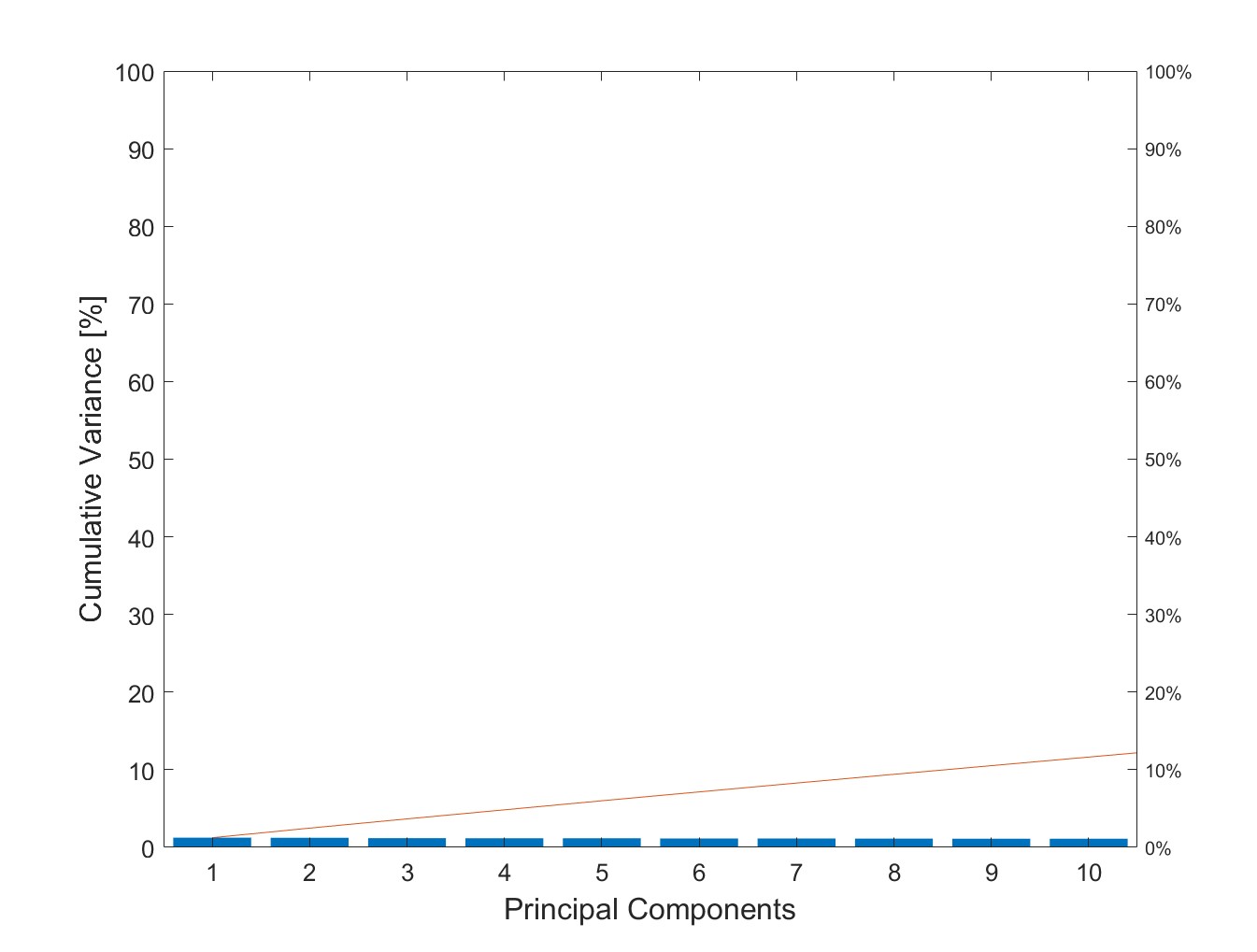} % left figure
		\caption{Bar plot of the principal components with a cumulative variance curve.}\label{rys:PCA1_1}
	\end{minipage}
	\begin{minipage}[b]{0.49\textwidth}
		\centering
		\includegraphics[width=0.98\textwidth]{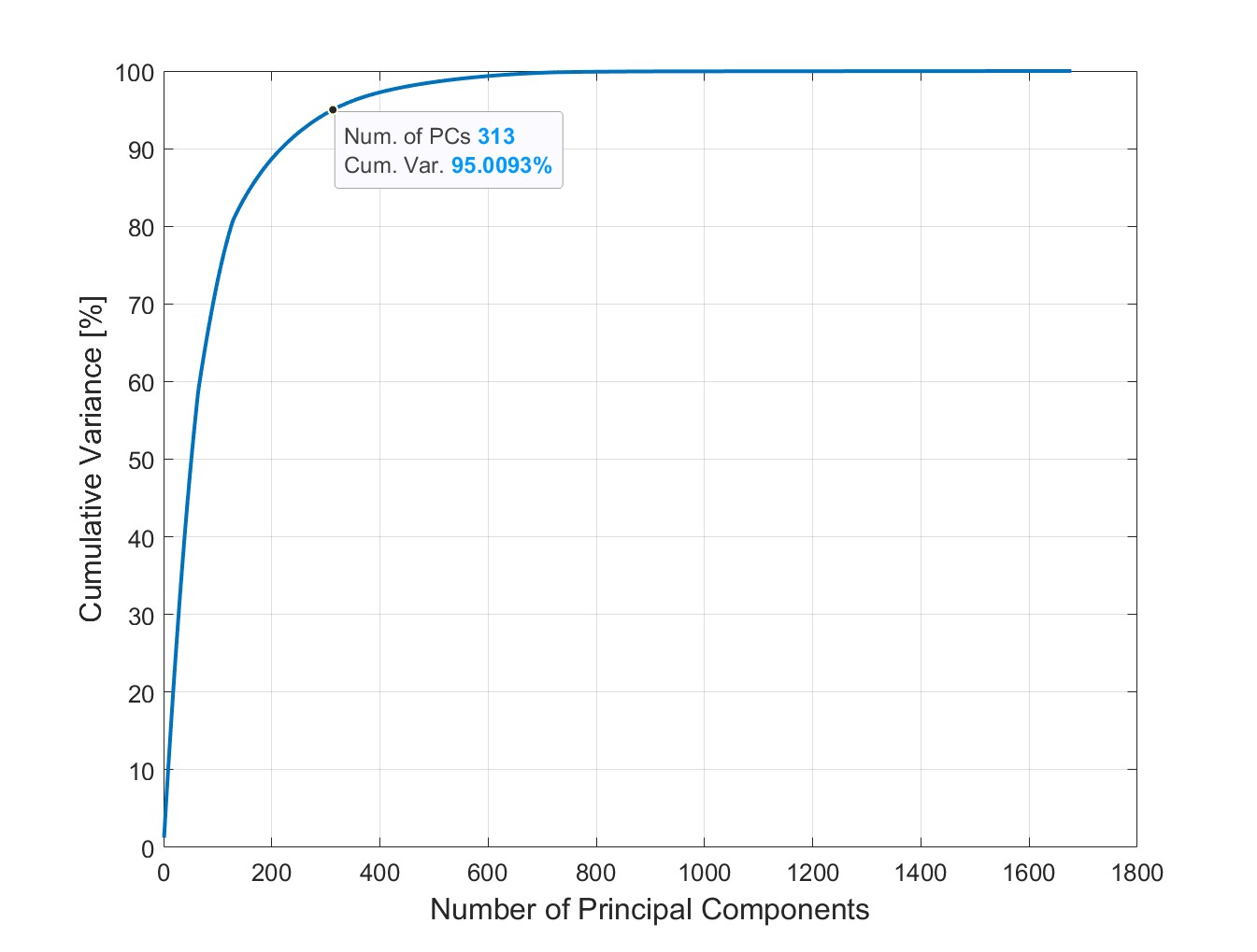} % right figure
		\caption{Cumulative variance based on the number of major components graph.}\label{rys:PCA2_2}
	\end{minipage}
\end{figure}

Based on the analysis of the graphs depicting the relationship between cumulative variance and the number of principal components (\ref{rys:PCA1_1}, \ref{rys:PCA2_2}), it can be concluded that each of the main components contributes a small but even value to the cumulative variance of the data, amounting to about 1\% on average. By utilizing 314 principal components, a cumulative variance of 95\% was achieved, thereby reducing the number of parameters from 1678 features.
\begin{figure}[!h]
\centering\includegraphics[width=0.5\textwidth]{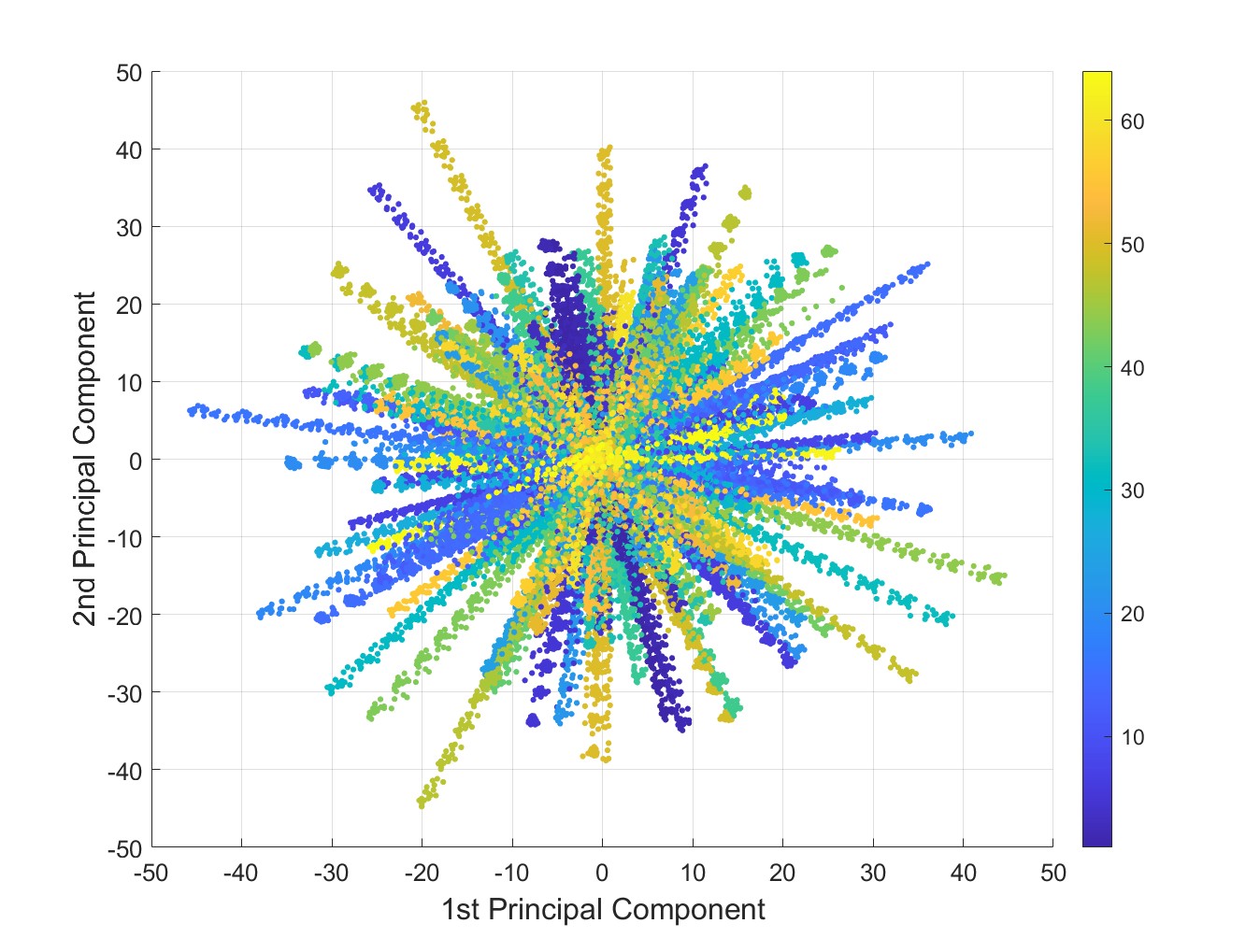}
\caption{Visualization of the first two principal components.}  \label{rys:PCA3_3}
\end{figure}
The visualization of the data projected onto the first two principal components suggests a high degree of class separability. The graph \ref{rys:PCA3_3} demonstrates that class separation improves as the distance from the origin of the coordinate system increases.

The Principal Component Analysis (PCA) algorithm can be effectively implemented through Singular Value Decomposition (SVD) \cite{brunton_svd_2019}.

\section{Support Vector Machine}
The Support Vector Machine (SVM) \cite{bishop2006pattern} is a widely utilized and highly effective supervised learning algorithm in machine learning. Its primary objective is to determine an optimal hyperplane within the feature space that maximally separates data points belonging to distinct classes. The hyperplane is defined such that the margin between the closest data points from opposing classes is maximized. These pivotal data points, referred to as support vectors, play a crucial role in shaping the classification boundary.
\subsection{SVM Parameters}
A critical factor in optimizing the performance of the Support Vector Machine (SVM) is the meticulous selection of parameters, as they directly impact model accuracy, generalization ability, and computational efficiency.
\subsubsection{Regularization Parameter \( C \)}
The regularization parameter CC regulates the balance between model complexity and its generalization capability, influencing the trade-off between minimizing training error and preventing overfitting..
\begin{itemize}
\item \textbf{High C}: Emphasizes accurate classification of training data,            resulting in a narrower margin. However, this increases the risk of overfitting, potentially reducing the model's generalization ability.
\item \textbf{Low C}: Promotes a wider margin by permitting a higher degree of misclassification during training, enhancing generalization, especially in noisy datasets. Nonetheless, this may lead to underfitting.
\end{itemize}
\subsubsection{Kernel Function Parameters}
The choice of kernel function significantly impacts SVM performance, as it determines the transformation of the input space and the parameters used in classification.
\begin{table}[h!]
\centering
\caption{SVM Kernel Functions and Parameters}
\label{tab:kernels}
\begin{tabular}{|c|c|p{4cm}|}
  \hline
  \textbf{Kernel} & \textbf{Parameters} & \textbf{Description} \\ \hline
  Linear & - & Assumes linear \\
         &   & separability of data \\
         &   & and does not introduce \\
         &   & additional complexity. \\ \hline
  Gaussian/RBF & \(\gamma\) & Defines the influence \\
              &           & of individual training \\
              &           & points; higher values \\
              &           & focus on local patterns, \\
              &           & whereas lower values \\
              &           & promote a more \\
              &           & generalized structure. \\ \hline
  Polynomial & \(d\) & Specifies the \\
            &      & polynomial degree, \\
            &      & affecting the \\
            &      & complexity of the \\
            &      & decision boundary. \\ \hline
  Sigmoid & \(\alpha, c\) & \(\alpha\): Determines \\
          &              & the slope of the \\
          &              & function; \(c\): Controls \\
          &              & the decision boundary \\
          &              & shift. \\ \hline
  \end{tabular}
\end{table}
\subsubsection{Multi-Class Classification}
Since SVM is inherently a binary classifier, multi-class classification is achieved using strategies such as One-vs-All and One-vs-One.
\paragraph{One-vs-All (OvA)}
This approach constructs \( K \) binary classifiers, where each classifier distinguishes one class from all others.
\begin{itemize}
    \item \textbf{Advantages}: Simple to implement, computationally efficient for large datasets, and provides straightforward interpretability.
    \item \textbf{Disadvantages}: May struggle with imbalanced datasets and can lead to ambiguity when multiple classifiers assign the same instance to their respective classes.
\end{itemize}
\paragraph{One-vs-One (OvO)}
This method constructs \( \frac{K(K-1)}{2} \) binary classifiers, where each classifier is trained on a distinct pair of classes.
\begin{itemize}
    \item \textbf{Advantages}: Typically achieves higher accuracy and robustness due to majority voting.
    \item \textbf{Disadvantages}: Computationally expensive due to the large number of classifiers and more challenging to interpret compared to the One-vs-All approach.
\end{itemize}
\subsection{Model parameters optimization}
During the design of the model, five time cross-validation was used. This procedure involved dividing the data into five groups, with one group serving as the test set and the others as the training set in each iteration, averaging the results and limiting the impact of random splitting of the data on the final model evaluation.
To maximize the model's accuracy, hyperparameter optimization \cite{MathWorksClassificationLearnerApp,MathWorksStatisticsToolbox} was used, which automatically searched the space of possible parameter configurations,
selecting those that provided the best accuracy for model prediction. The optimization of SVM model parameters was done in 30 iterations using a Bayesian optimization algorithm. This algorithm iteratively updates assumptions about the hyperparameter space to find optimal values that minimize the error function.
To identify the most promising points in the parameter space, the "Expected Improvement per Second Plus" (EIPS) acquisition function was used. This function takes into account both expected improvements and training time, allowing for more efficient allocation of computational resources, especially with large datasets, and helps reduce the number of required iterations.

The chart below shows the relationship between the minimum classification error and the number of iterations, as well as the results of model parameter optimization.
\begin{figure}[!h]
    \centering
    \includegraphics[width=1\linewidth]{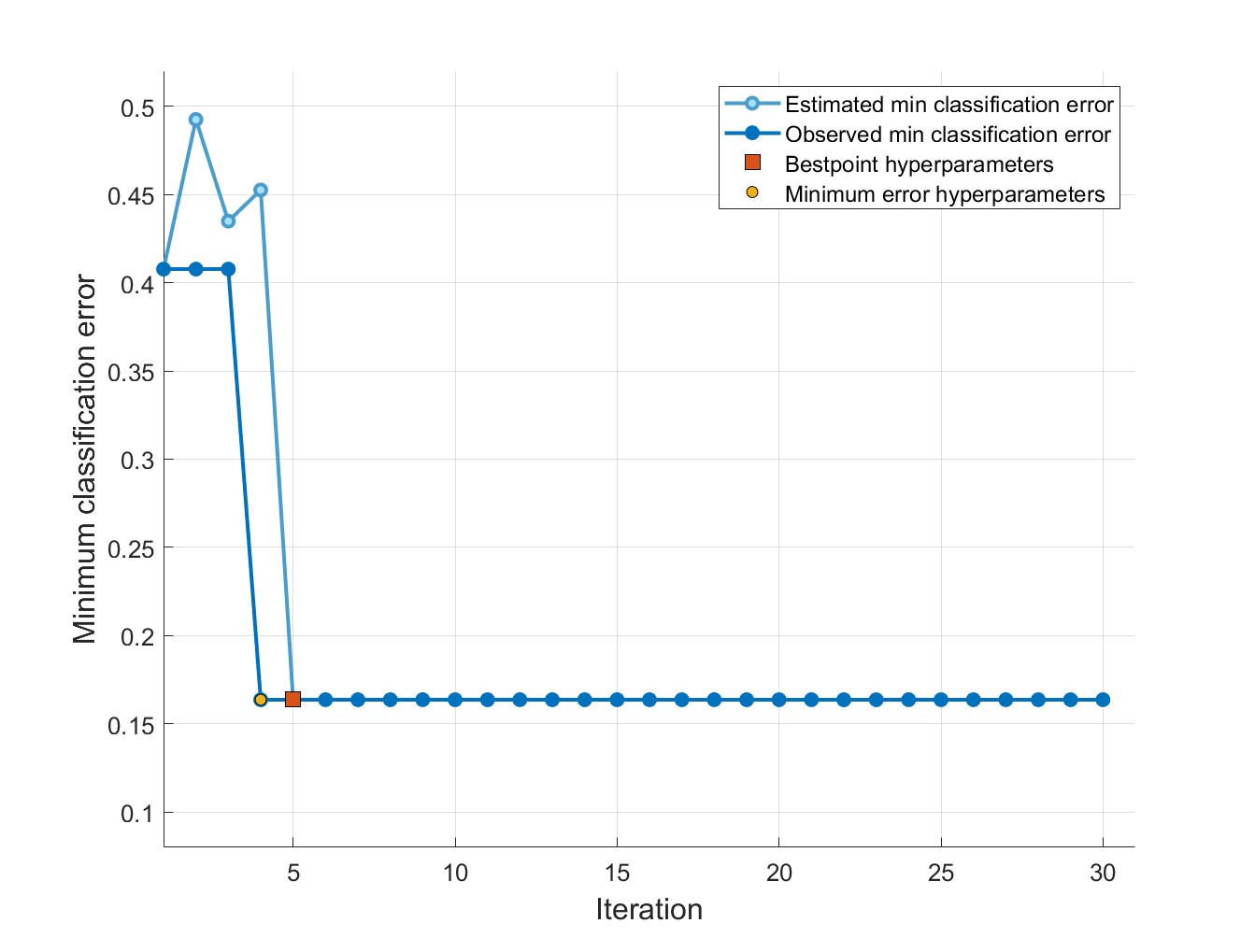}
    \caption{The graph of the minimum classification error depending on the number of iterations.}
    \label{fig:optimization}
\end{figure}

The parameters obtained after optimizing the SVM are shown below.
\begin{table}[h!]
    \centering
    \caption{SVM Model Parameters}
    \begin{tabular}{|l|l|}
        \hline
        \textbf{Parameter}             & \textbf{Value}        \\ \hline
        Kernel                         & Quadratic             \\ \hline
        Box Constraint (C parameter)  & 0.09                  \\ \hline
        Kernel Scale                   & 1             \\ \hline
        Multiclass Coding              & One-vs-All            \\ \hline
        Data Standardization           & No                    \\ \hline
    \end{tabular}
    \label{tab:svm_parameters}
\end{table}

\subsection{Training results}
A confusion matrix was generated for the trained model. The accuracy of the model on the validation set was also calculated, which amounted to 94.8\%.

\begin{figure}[!h]
    \centering
    \includegraphics[width=1\linewidth]{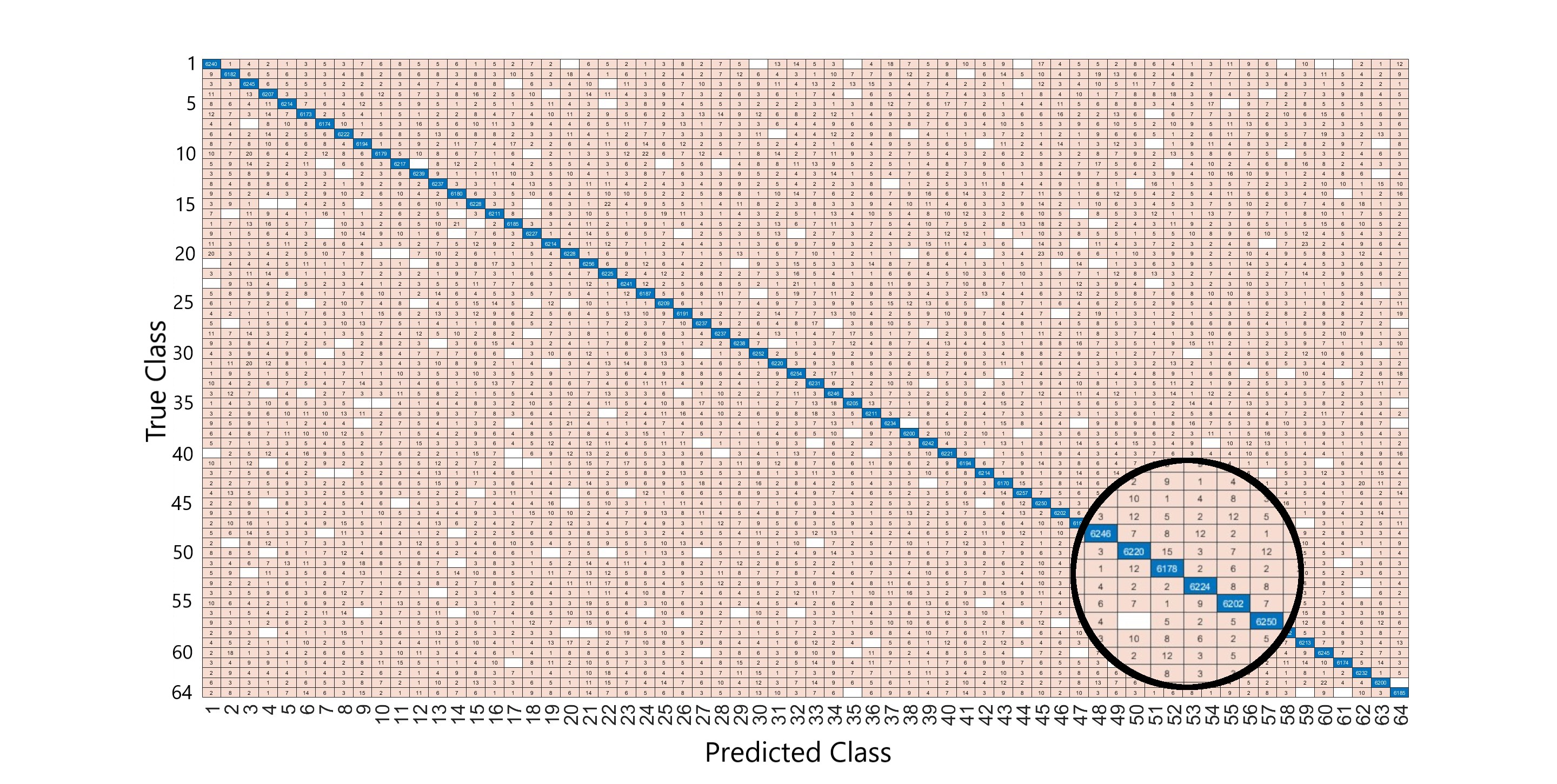}
    \caption{Confusion matrix of the SVM model.}
    \label{fig:conf_matrix}
\end{figure}

\begin{table}[ht]
    \centering
    \caption{Model accuracy on the validation set.}
    \begin{tabular}{|l|c|}
    \hline
    \textbf{Accuracy on the validation set} & \textbf{Value} \\
    \hline
    Accuracy  & 94.8\% \\
    \hline
    \end{tabular}
\end{table}

The accuracy achieved considering as many as 64 classes is a satisfying result that confirms the effectiveness of the parameter optimization carried out. It's a good sign before conducting tests on the 4G/5G simulator.

\section{Model testing}

To validate the effectiveness of the developed PRACH preamble detector and compare its performance with an analytical algorithm, simulations were conducted using the LTE/5G simulator previously employed for database generation. These simulations focused on an early access procedure scenario and were performed for both the developed detector and the analytical algorithm. The experiments covered various propagation conditions, allowing for a comprehensive assessment of both solutions under diverse scenarios.

\subsection{Simulation Parameters}
The simulation parameters for the scenario were as follows:
\begin{itemize}
    \item The transmission Signal-to-Noise Ratio (SNR) varied from -20 dB to 20 dB,
    \item The simulation duration was set to 10 seconds,
    \item One physical receiver antennas were used,
    \item A single User Equipment (UE) was present in the simulation,
    \item A fixed Timing Advance (TA) was applied,
    \item The PRACH sequence length was generated for the long format 0,
\end{itemize}

\subsection{Propagation Channel Models}
The testing was conducted on the following propagation channel models \cite{3GPP-LTE,MathWorksPropagationChannel,3GPP-TS-38.101-4}:

\begin{itemize}
    \item 	Additive White Gaussian Noise (AWGN)
    \item 	Extended Pedestrian A (EPA)
    \item 	Extended Vehicular A (EVA)
    \item 	Extended Typical Urban (ETU)
    \item 	TDLC300
    \item 	TDLD30
\end{itemize}

By analyzing the performance of the PRACH preamble detector under these channel conditions, its robustness and adaptability to different real-world scenarios were evaluated.

\section{Results}
According to 3GPP standards \cite{3GPP-LTE}, the probability of a false alarm should not exceed 0.1\%, while the probability of preamble detection should reach at least 99\%. Ensuring such high precision is crucial for the proper functioning of LTE and 5G networks.

The following charts illustrate the comparison of FAR and MDR results between the SVM detector and a legacy algorithm. The results encompass all mentioned propagation channels. The MDR and FAR values on the charts are identical, as each instance of preamble misclassification was simultaneously interpreted by the SVM detector as an assignment of a preamble other than the one actually sent (FA) and as a failure to correctly detect the preamble that was indeed transmitted (MD).

\begin{figure}[!h]
\centering
\includegraphics[width=0.5\textwidth]{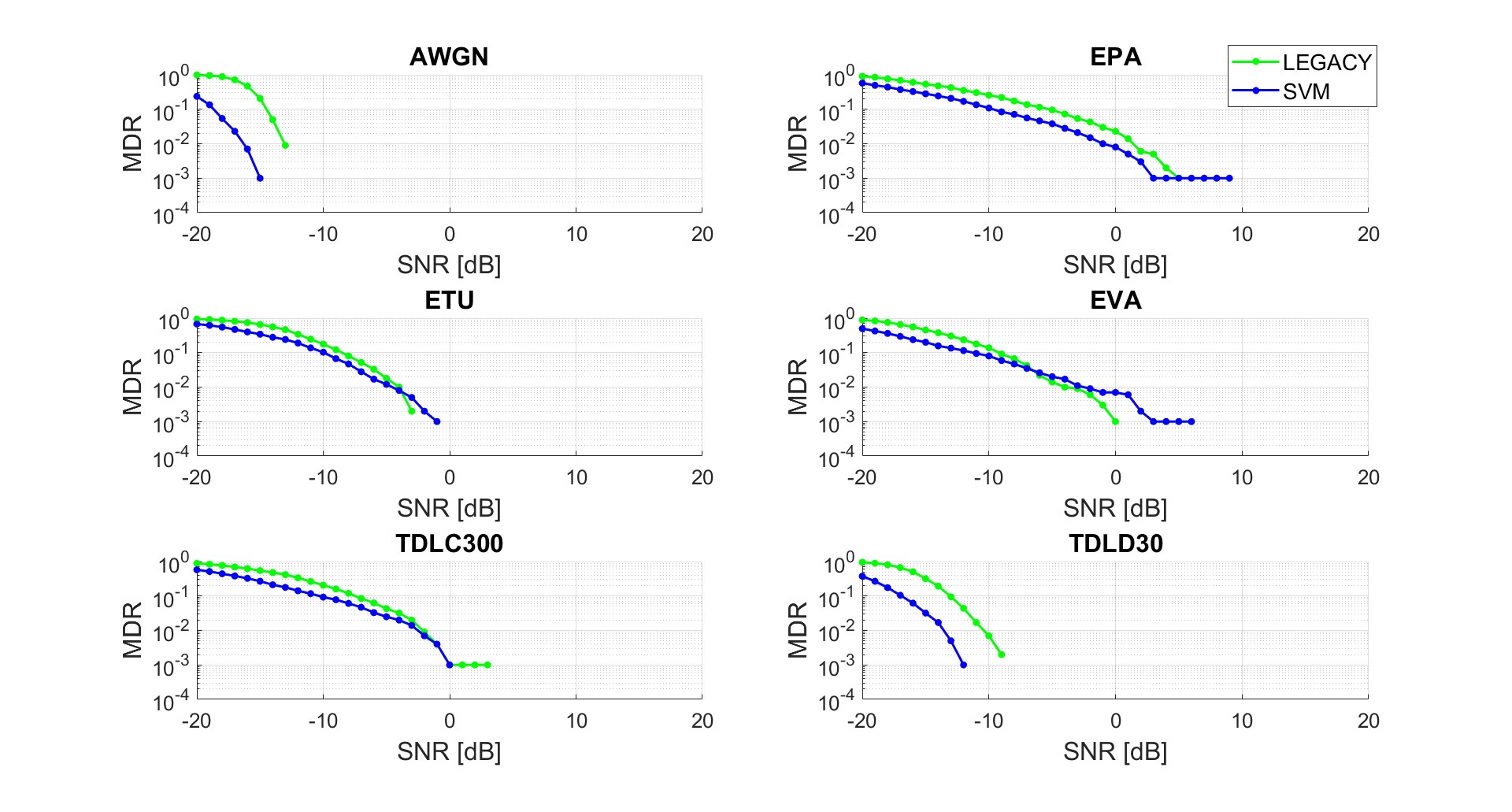}
\caption{MDR results comparision for legacy algorithm and SVM detector.}
\label{fig_sim}
\end{figure}

\begin{figure}[!h]
\centering
\includegraphics[width=0.5\textwidth]{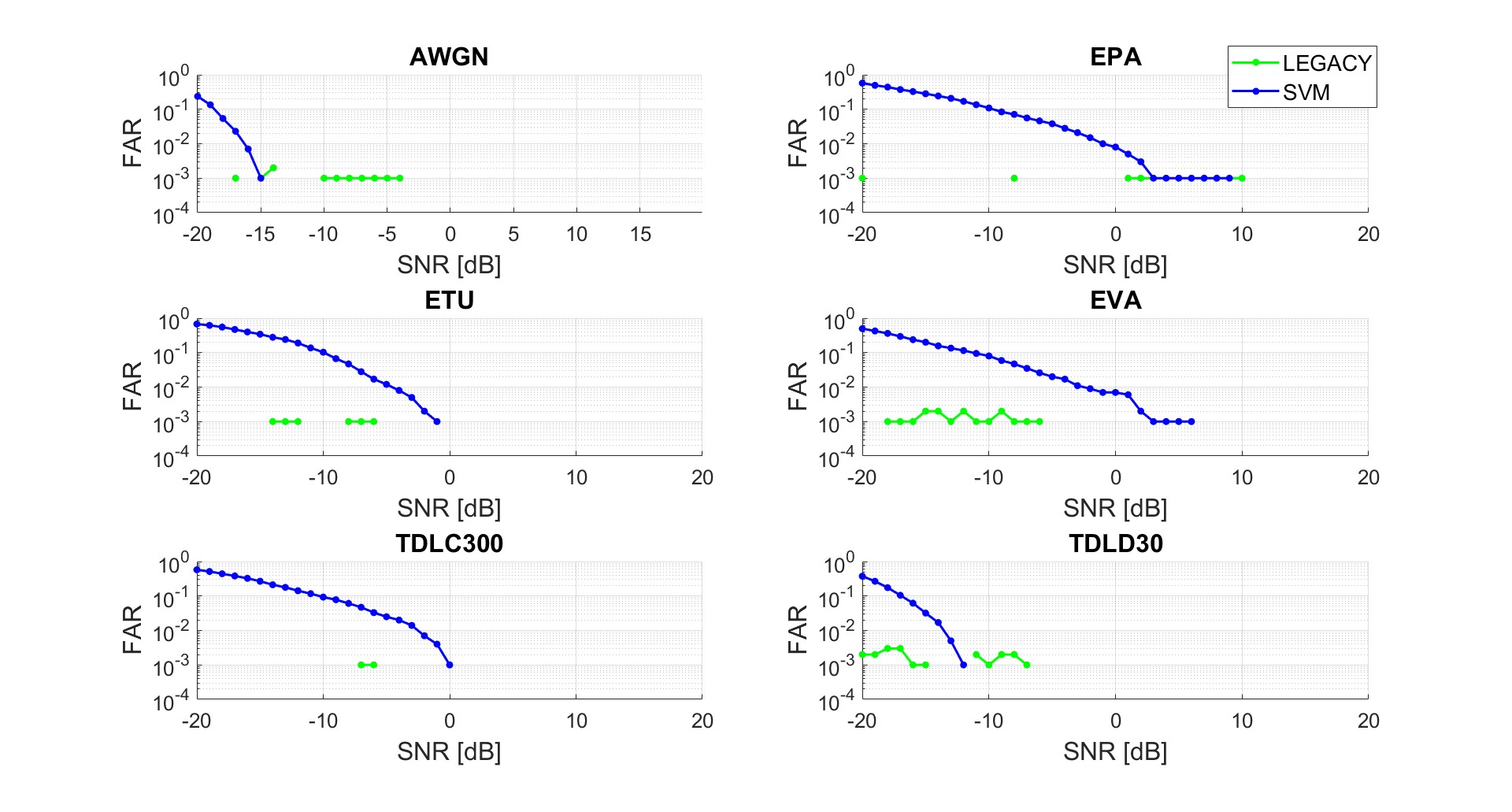}
\caption{FAR results comparision for legacy algorithm and SVM detector.}
\label{fig_sim}
\end{figure}

The analytical algorithm achieved a reference level of MDR at -13 dB for the AWGN channel, 2 dB for the EPA channel, -3 dB for the EVA channel, -4 dB for the ETU, -2 dB for the TDLC300, and -11 dB for the TDLD30. False alarms occurred only at selected SNR points; however, their number did not exceed 0.3\% across all analyzed propagation channels. The worst FAR result was obtained for the TDLD30 channel. The FAR results for the analytical algorithm are uneven, with false alarms appearing in random ranges.

The SVM-based detector achieved a reference level of the MDR indicator for an SNR level of: -16 dB in the AWGN channel, -1 dB for EPA, -3 dB for EVA, -5 dB for ETU, -3 dB for TDLC300, and -14 dB in the TDLD30 channel. For the FAR indicator, the detector attained a value of 0.1\% for SNR levels of: -15 dB in the AWGN channel, 3 dB in the EPA channel, 3 dB in the EVA channel, and -1 dB in the ETU channel.

\section{Conclusion}
When comparing the results of the SVM detector with those of the legacy algorithm, it can be observed that for the AWGN channel, the developed detector achieved the reference MDR value at an SNR level that was 3 dB better, for the EPA channel by 3 dB, for ETU and TDLC300 by 1 dB, and for TDLD30 it was 3 dB. The only channel where the SVM-based detector performed worse than the legacy algorithm was the EVA channel, where the reference MDR level was achieved 1 dB later than in the classical approach. Such differences in SNR levels are very significant in telecommunications systems and can greatly affect the system's operational efficiency.

As outlined in the previous section, the assumption of not disregarding the noise class results in the algorithm always producing a detection, which contributes to a high false alarm (FA) rate. In this context, the SVM algorithm has the potential to serve as a complementary approach to the classical solution, aiding in the reduction of the missed detection (MD) rate. Future research should focus on defining noise as a distinct detection class to further enhance the algorithm’s performance.

The obtained results are promising, demonstrating the appropriate selection of parameters in the optimization process and highlighting the potential for further enhancements of the developed solution. Future research directions may involve expanding the training dataset and training the model on signals from rapidly moving objects. Additionally, after such an expansion, an in-depth examination of the optimal parameter selection for the SVM algorithm should be conducted.


% \section*{Acknowledgment}


%\bibliographystyle{IEEEtran}

\begin{thebibliography}{1}

\bibitem{3GPP-TR-38.802} 3rd Generation Partnership Project (3GPP), "Study on New Radio (NR) Access Technology; Physical Layer Aspects," 3GPP Technical Report TR 38.802, vol. V14.2.0, Mar. 2017.

\bibitem{3GPP-TR-38.801} 3rd Generation Partnership Project (3GPP), "Study on New Radio Access Technology; Radio Access Architecture and Interfaces," 3GPP Technical Report TR 38.801, vol. V14.0.0, Mar. 2017.

\bibitem{Guel2024} D. Guel, A. Kabore, and D. Bassole, "5G NR PRACH Detection with Convolutional Neural Networks (CNN): Overcoming Cell Interference Challenges," arXiv preprint, 2024. [Online]. Available: \url{https://arxiv.org/abs/2408.11659}.

\bibitem{Modina2019} N. Modina, R. Ferrari, and M. Magarini, "A Machine Learning-Based Design of PRACH Receiver in 5G," in Proceedings of the 2nd International Workshop on Recent Advances in Cellular Technologies and 5G for IoT Environments (RACT-5G-IoT), Leuven, Belgium, Apr. 2019, pp. 3. [Online]. Available: \url{https://re.public.polimi.it/retrieve/handle/11311/1124713/477363/RACT-5G-IoT_3_6791_&_6834.pdf}

\bibitem{Singh2024} R. Singh, A. K. Yerrapragada, and R. K. Ganti, "A Machine Learning based Hybrid Receiver for 5G NR PRACH," arXiv preprint, 2024. [Online]. Available: \url{https://arxiv.org/abs/2411.08919}.

\bibitem{Zehra2022} S. S. Zehra, M. Magarini, and R. Qureshi et al., "Proactive approach for preamble detection in 5G-NR PRACH using supervised machine learning and ensemble model," Scientific Reports, vol. 12, p. 8378, 2022. [Online]. Available: \url{https://doi.org/10.1038/s41598-022-12349-4}.

\bibitem{Li2024} R. Li, Y. Chu, J. Xue, J. Sun, and S. Chatzinotas, "A Deep Learning Approach for Universal NPRACH Detection with Inter-Cell Interference," IEEE, 2024. [Online]. Available: \url{https://www.nature.com/articles/s41598-022-12349-4}

\bibitem{ZC-seq} J. G. Andrews, "A Primer on Zadoff-Chu Sequences," Nov. 2022. [Online]. Available: \url{https://ar5iv.labs.arxiv.org/html/2211.05702}

\bibitem{MathWorksRACH} MathWorks, "Random Access Channel." [Online]. Available: \url{https://www.mathworks.com/help/lte/ug/random-access-channel.html}

\bibitem{ShareTechNote} ShareTechNote, "5G/NR - Initial Access/RACH." [Online]. Available: \url{https://www.sharetechnote.com/html/5G/5G_RACH.html}

\bibitem{3GPP-TS-38.211} 3rd Generation Partnership Project (3GPP), "NR; Physical channels and modulation (Release 18)," 3GPP Technical Specification TS 38.211, vol. V18.4.0, Sep. 2024.

\bibitem{3GPP-TS-38.101-4} 3rd Generation Partnership Project (3GPP), "NR; User Equipment (UE) radio transmission and reception; Part 4: Performance requirements (Release 18)," 3GPP Technical Specification TS 38.101-4, vol. V18.5.0, Sep. 2024.

\bibitem{keysight-5G} Keysight Technologies, "Understanding the 5G NR Physical Layer." [Online]. Available: \url{https://www.keysight.com/us/en/assets/9921-03326/training-materials/Understanding-the-5G-NR-Physical-Layer.pdf}

\bibitem{Lin2019-5G} X. Lin, J. Li, R. Baldemair, et al., "5G New Radio: Unveiling the Essentials of the Next Generation Wireless Access Technology," arXiv, 2018. [Online]. Available: \url{https://arxiv.org/abs/1806.06898}

\bibitem{Jolliffe-Cadima-PCA} I. T. Jolliffe and J. Cadima, "Principal Component Analysis: A Review and Recent Developments," Philosophical Transactions of the Royal Society A, vol. 374:20150202, 2016. doi: 10.1098/rsta.2015.0202.

\bibitem{MathWorksPCA} MathWorks, "PCA." [Online]. Available: \url{https://www.mathworks.com/help/stats/pca.html}

\bibitem{eriksen2023data} T. Eriksen and N.u. Rehman, "Data-driven nonstationary signal decomposition approaches: a comparative analysis," Scientific Reports, vol. 13, 2023. doi: 10.1038/s41598-023-28390-w.

\bibitem{brunton_svd_2019} S. L. Brunton and J. N. Kutz, "Singular Value Decomposition (SVD)," in Data-Driven Science and Engineering: Machine Learning, Dynamical Systems, and Control, Cambridge University Press, 2019, pp. 3-46.

\bibitem{bishop2006pattern} C. M. Bishop, "Pattern Recognition and Machine Learning," Springer, New York, 2006. Chapter 7: Sparse Kernel Machines.

\bibitem{MathWorksClassificationLearnerApp} MathWorks, "Classification Learner App." [Online]. Available: \url{https://www.mathworks.com/help/stats/classification-learner-app.html}

\bibitem{MathWorksStatisticsToolbox} MathWorks, "Statistics and Machine Learning Toolbox." [Online]. Available: \url{https://www.mathworks.com/help/stats/index.html?s_tid=CRUX_lftnav}

\bibitem{3GPP-LTE} 3rd Generation Partnership Project (3GPP), "Technical Specification Group Radio Access Network; Evolved Universal Terrestrial Radio Access (E-UTRA); Base Station (BS) radio transmission and reception, Release 18," 3GPP Technical Specification TS 36.104, vol. V18.5.0, Mar. 2024. [Online]. Available: \url{https://portal.3gpp.org/desktopmodules/Specifications/SpecificationDetails.aspx?specificationId=2412}

\bibitem{MathWorksPropagationChannel} MathWorks, "Propagation Channel Models." [Online]. Available: \url{https://www.mathworks.com/help/lte/ug/propagation-channel-models.html}

\end{thebibliography}

\end{document}